\documentclass[%
 reprint,
superscriptaddress,
 amsmath,amssymb,
 aps,
]{revtex4-2}

\usepackage{graphicx}
\usepackage{dcolumn}
\usepackage{bm}
\usepackage{multirow}
\usepackage{color}

\newcommand{\eg}{$e_{g}$}
\newcommand{\tg}{$t_{2g}$}
\newcommand{\dzt}{$d_{z^2}$}
\newcommand{\dxtyt}{$d_{x^2-y^2}$}
\newcommand{\dxy}{$d_{xy}$}
\newcommand{\dxz}{$d_{xz}$}
\newcommand{\dyz}{$d_{yz}$}

\newcommand{\mub}{$\mu_{\rm B}$}
\newcommand{\ef}{$E_{\rm F}$}

\begin{document}

\title{Electronic reconstruction and charge transfer in strained Sr$_2$CoIrO$_6$ double perovskite}

\author{Jiongyao Wu}
\affiliation{Department of Physics and Center for Nanointegration Duisburg-Essen (CENIDE), University of Duisburg-Essen, Lotharstr. 1, D-47057 Duisburg, Germany}
\author{Marcel Ney}
\affiliation{Department of Physics and Center for Nanointegration Duisburg-Essen (CENIDE), University of Duisburg-Essen, Lotharstr. 1, D-47057 Duisburg, Germany}
\author{Sebastian Esser}
\affiliation{Experimental Physics $VI$, Center for Electronic Correlations and Magnetism, Augsburg University, D-86159 Augsburg, Germany}
\author{Vijaya Begum}
\affiliation{Department of Physics and Center for Nanointegration Duisburg-Essen (CENIDE), University of Duisburg-Essen, Lotharstr. 1, D-47057 Duisburg, Germany}
\author{G\"unther Prinz}
\affiliation{Department of Physics and Center for Nanointegration Duisburg-Essen (CENIDE), University of Duisburg-Essen, Lotharstr. 1, D-47057 Duisburg, Germany}
\author{Axel Lorke}
\affiliation{Department of Physics and Center for Nanointegration Duisburg-Essen (CENIDE), University of Duisburg-Essen, Lotharstr. 1, D-47057 Duisburg, Germany}
\author{Philipp Gegenwart}
\affiliation{Experimental Physics $VI$, Center for Electronic Correlations and Magnetism, Augsburg University, D-86159 Augsburg, Germany}
\author{Rossitza Pentcheva}
\email{Rossitza.Pentcheva@uni-due.de}
\affiliation{Department of Physics and Center for Nanointegration Duisburg-Essen (CENIDE), University of Duisburg-Essen, Lotharstr. 1, D-47057 Duisburg, Germany}
\pacs{}
\date{\today}

\begin{abstract}
	The electronic,  magnetic and  optical properties  of the double perovskite Sr$_2$CoIrO$_6$ (SCIO) under biaxial strain are explored in the framework of density functional theory (DFT) including a Hubbard $U$ term and spin-orbit coupling (SOC) in combination with absorption spectroscopy measurements on epitaxial thin films. While the end member SrIrO$_3$ is a semimetal with a quenched spin and orbital moment and bulk SrCoO$_3$ is a ferromagnetic (FM) metal with spin and orbital moment of 2.50 and 0.13 $\mu_{B}$, respectively, the double perovskite SCIO emerges as an antiferromagnetic Mott insulator with antiparallel alignment of  Co, Ir planes along the [110]-direction. Co exhibits a spin and enhanced orbital moment of $\sim 2.35-2.45$ and $0.31-$0.45 $\mu_{B}$, respectively. Most remarkably, Ir acquires a significant spin and orbital moment of 1.21-1.25 and 0.13 $\mu_{B}$, respectively. Analysis of the orbital occupation indicates an electronic reconstruction due to a substantial charge transfer from minority to majority spin states in Ir and from Ir to Co, signaling an Ir$^{4+\delta}$, Co$^{4-\delta}$ configuration. Biaxial strain, varied from -~1.02$\%$ ($a_{\rm NdGaO_3}$) through 0\% ($a_{\rm SrTiO_3}$) to 1.53$\%$ ($a_{\rm GdScO_3}$), influences in partcular the orbital polarization of the $t_{2g}$ states and leads to a nonmonotonic change of the band gap between 163 and 235 meV. The absorption coefficient reveals a two plateau fearure due to transitions from the valence to the lower lying narrow $t_{2g}$ and the higher lying broader $e_{g}$ bands.  
Inclusion of many body effects, in particular, excitonic effects by solving the Bethe-Salpeter equation (BSE), increases the band gap by $\sim0.2$ and improves the agreement with the measured spectrum concerning the position of the second peak at $\sim 2.6$ eV.
\end{abstract}
\maketitle


\section{\label{sec:Introduction}Introduction}

The interplay of charge, spin, lattice and orbital degrees of freedom in transition metal oxides (TMO) ~\cite{Tokura2000} gives rise to a  rich functionality. The double perovskites A$_2$BB$^{'}$O$_6$ are a special class of TMO that allow to combine elements with different characteristics at the B and the B$^{'}$ sites and thus achieve novel behavior~\cite{Karppinen},  that can vary between ferromagnetic halfmetallic with high $T_{\rm C}$ as e.g. in Sr$_2$FeMoO$_6$ to ferromagnetic insulating as in La$_2$CoMnO$_6$~\cite{Fournier} and Ba$_2$CuOsO$_6$~\cite{Yamaura}.  A variety of Ir-based double perovskites have been considered recently owing to their interesting electronic and magnetic properties, resulting from the interplay of spin-orbit coupling (SOC), electronic correlation and structural distortions: 
For example, La$_2$MgIrO$_6$ and La$_2$ZnIrO$_6$ are reported as spin-orbit magnetic Mott insulators~\cite{Mandrus}. Selecting $3d$ and $5d$ transition metal ions allows to combine the enhanced electronic correlations of the former with the strong SOC of the latter. Sr$_2$CoIrO$_6$ (SCIO) is an interesting example in this context. 

  The perovskite end member SrCoO$_3$ (SCO) shows ferromagnetic (FM) behavior at temperatures below $T_{\rm C}$ = 280–305~K \cite{Tokura}. Recent DFT+$U$ studies suggested that the tetragonal $P4/mbm$ phase with moderate Jahn-Teller distortions is preferred over the $Pm\bar{3}m$ phase~\cite{Cazorla}. Both experiments and theoretical calculations reveal intermediate spin (IS) state with a magnetic moment of $\sim$2.6 $\mu_{B}$ \cite{Tokura, Abbate, Ernst}. Epitaxial strain is predicted to drive  a transition from  a ferromagnetic-metallic (FM-M)  to an antiferromagnetic (AFM) insulating and ferroelectric phase~\cite{Rabe}. 
  
  In the other end member, SrIrO$_3$ (SIO), strong SOC leads to a splitting of the Ir-$t_{2g}$ orbitals into completely filled $j_{\rm eff}=3/2$ band and half-filled $j_{\rm eff}=1/2$. While SIO exhibits a quenched spin and orbital magnetic moment and topological crystalline semimetalic character  \cite{Kee, KeeH, Zhao, Arnott, Noh} in space group $Pbnm$~\cite{Rondinelli}, in other iridates such as the Ruddlesden-Popper (RP) Sr$_2$IrO$_4$ with Ir$^{4+}$ (5d$^5$)  the interplay of electronic correlations and SOC opens a gap giving rise to a $j_{\rm eff}=1/2$ Mott insulator~\cite{Balents, Rotenberg, Arima, Franchini, Noh, Khaliullin, McMorrow}.

Previous reports on the bulk double perovskite SCIO~\cite{Ehrenberg, EhrenbergH, Gegenwart, Tjeng} suggested an Ir$^{5+}$, Co$^{3+}$ configuration, though there is a controversy on the magnetic state of Ir \cite{EhrenbergH, Tjeng}: While the former study reports a nearly quenched Ir total magnetic moment of 0.039 \mub, the latter proposes the presence of a small fraction ($\sim 10~\%$) of Ir$^{6+}$ ($s=3/2$) to explain a finite Ir magnetic moment.  Recently, thin films of SCIO were grown on different substrates using metalorganic aerosol deposition (MAD)~\cite{Gegenwart}. Polarization-dependent x-ray absorption (XAS) measurements point towards strain dependent changes in the orbital configuration and magnetocrystalline anisotropy. This goes hand in hand with a sign change of the magnetoresistance at low temperatures that could be explained by a quantitative model including anisotropic magnetoresistance and weak antilocalization effects.

\begin{figure}[!htp]
\includegraphics[width=0.45\textwidth]{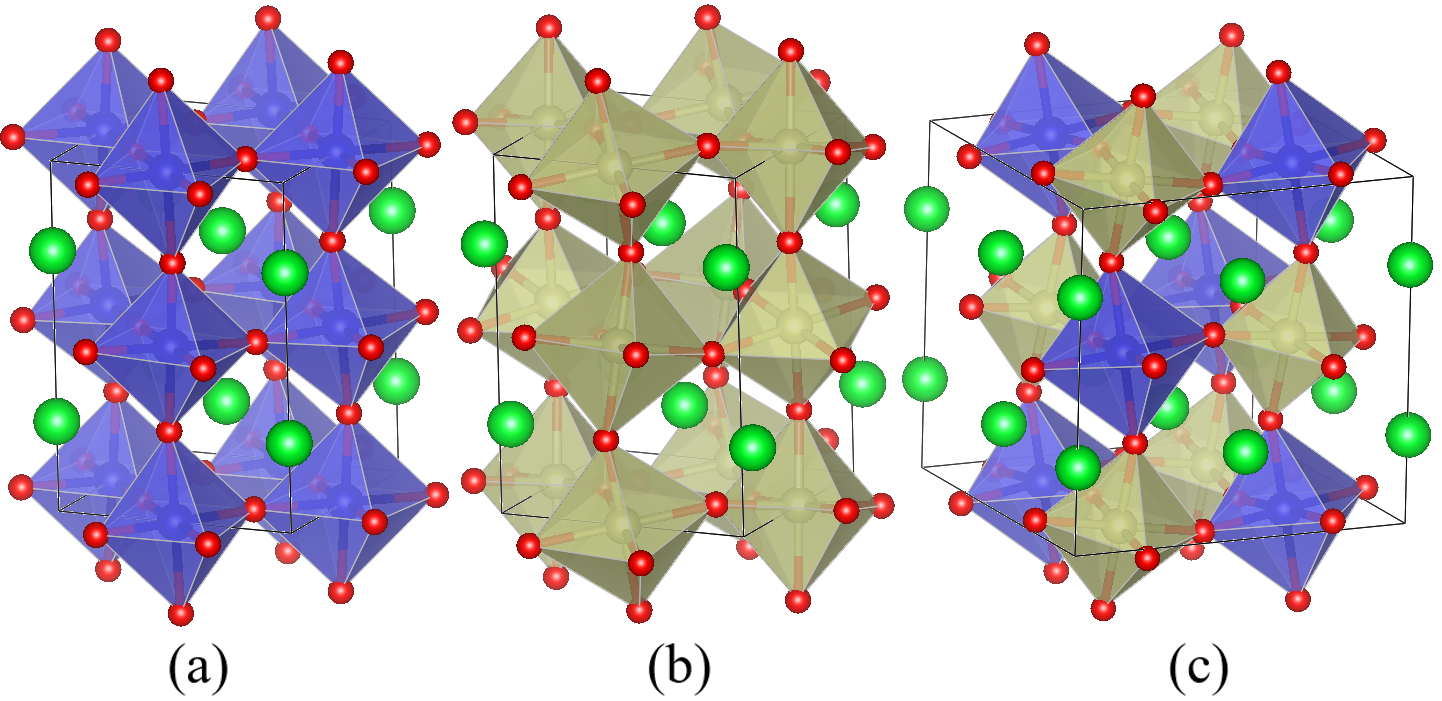}%
\caption{\label{fig:structure} Structure of (a) SrCoO$_3$, (b) SrIrO$_3$ and (c) double perovskite Sr$_2$CoIrO$_6$. Sr, Co, Ir and O ions are shown in green, yellow, blue and red, respectively. Note that the DP is modeled in a 40-atom unit cell, rotated by 45$^{\circ}$ and twice the size of the ones of the end members.}
\end{figure}
  
Motivated by these results, we report here a comprehensive DFT+$U$ investigation including  spin-orbit effects of the structural, electronic and magnetic properties of the double perovskite (DP)  SCIO as a function of strain in comparison  to the end members SCO and SIO.  Our results confirm the semimetallic ground state of SrIrO$_3$. SrCoO$_3$ has a ferromagnetic metallic ground state in close proximity to an AFM insulating state, whereas SCIO emerges as an AFM insulator. The relaxed structures of SCO, SIO and SCIO, shown in Fig.~\ref{fig:structure}, exhibit  substantial octahedral rotations and tilts. In contrast to the Co$^{4+}$ and Ir$^{4+}$ electron configuration in the end members SCO and SIO, detailed analysis indicates a noticeable charge transfer from Ir to Co resulting in  Co$^{4-\delta}$ in the  intermediate-spin state (IS) and an Ir$^{4+\delta}$ configuration in the double perovskite SCIO, reminiscent of the charge transfer recently reported in SrMnO$_3$/SrIrO$_3$(001) superlattices~\cite{OkamotoNL}. We note that the DP itself can be regarded as a superlattice containing alternating SrCoO$_3$ and SrIrO$_3$ layers in the (111)-direction.  Furthermore, we explored the role of biaxial strain, by varying the lateral lattice constant from the one of NdGaO$_3$ (compressive), through SrTiO$_3$ (unstrained) to GdScO$_3$ (tensile). A subtle strain effect is observed in the orbital occupation and the band gap. Moreover, we have calculated the optical properties  within the independent-particle  (I.P.) picture, as well as by considering  quasiparticle effects within the $GW$ approximation (single shot $G_{0}W_{0}$)\cite{Hedin} and excitonic effects by solving the Bethe-Salpeter equation (BSE)\cite{Rohlfing, Albrecht}.
Changes of the in- and out-of-plane component of the real and imaginary part of the dielectric constant as a function of the lateral lattice parameter reflect the effect of strain. The calculated absorption coefficients are compared with transmission measurements on SCIO samples grown on different substrates using the MAD technique.

The paper is structured as follows: In Section~\ref{sec:CompDetail} and \ref{sec:ExpDetail} we describe the details of the theoretical modeling and experiments, respectively. In Section \ref{sec:strmag} and \ref{sec:elprop} the structural, electronic and magnetic properties of SCIO are compared to those of the end members. Furthermore, in \ref{sec:elpropCT} and \ref{sec:optical} a detailed analysis of the orbital occupation and optical properties as a function of strain is provided. The calculated absorption coefficients  are compared with transmission measurements on strained SCIO films. Finally, the results are summarized in Section~\ref{sec:Summary}.

\section{\label{sec:Meth}Methods}

\subsection{\label{sec:CompDetail}Computational details}

The density functional theory (DFT) calculations were performed with the VASP~\cite{Frthmuller, Joubert} code using the projector augmented wave (PAW) method~\cite{Blchl}. We chose the PBEsol~\cite{Burke, Perdew, PerdewJ} exchange-correlation functional, which is known to improve the structural description of solids as compared to PBE~\cite{PBE}. The following valence electron configuration was used: 4$s^{1}$3$d^{8}$ for Co, 6$s^{1}$5$d^{8}$ for Ir, 4$s^{2}$4$p^{6}$5$s^{2}$ for Sr and 2$s^{2}$2$p^{4}$ for O.
To account for static electronic correlations we  included a Hubbard $U$ term, using the Dudarev approach~\cite{Sutton}. We summarize here previously used $U$ values for related compounds, but  note that values applied or obtained with different codes are not directly comparable~\cite{Jian16,Oberhofer19}: $U_{\rm Co}=$5.44 eV and $U_{\rm Ir}=$2.77~eV, $J_{\rm Co}=$0.918 eV and $J_{\rm Ir}=$0.544 eV from ~\cite{Solovyev} were chosen in a previous study of SCIO~\cite{EhrenbergH} with the WIEN2k code\cite{wien2k}. For SCO the values previously used in VASP calculations range  from $U_{\rm Co}=$2.5 eV and $J_{\rm Co}=$1.0 eV~\cite{Rabe} to $U_{\rm Co}=6 eV$ ~\cite{Cazorla}. While element specific $U$ values in general depend on the local environment and oxidation state~\cite{Mason19}, for consistency and comparability we selected here the same $U$ values for the DP and the end members: $U=3.0$ eV for the Co ion, in order to describe correctly the FM metallic ground state of bulk SCO, as discussed below, as well as $U =1.37$ eV, $J=0.22$ eV for Ir, obtained from the constrained random phase approximation with the VASP code for SIO ~\cite{Franchini}. Additional calculations were performed with the Liechtenstein~\cite{Zaanen} approach, to ensure that the results of both approaches are consistent. In the following we discuss the ones obtained with the Dudarev approach.
  
  The wave functions were expanded in a plane-wave basis truncated at 600 eV. For SrCoO$_3$ and SrIrO$_3$ a 20-atom unit cell was used ($\sqrt{2}a$$\times$$\sqrt{2}a$$\times$$2c$), whereas a 40-atom unit cell was employed for  Sr$_2$CoIrO$_6$ in a $2a$$\times$$2a$$\times$$2c$ setup in order to model the cation and magnetic arrangement (see Fig.~\ref{fig:structure}). To sample the Brillouin zone, 8$\times$8$\times$6 $k$-points shifted off $\Gamma$ were generated using the Monkhorst-Pack scheme~\cite{Pack} for SrCoO$_3$ and SrIrO$_3$, and 5$\times$5$\times$5 $k$ points were used for the double perovskite Sr$_2$CoIrO$_6$. The convergence criterion for the self-consistent calculation is 10$^{-6}$ eV for the total-energy. A Fermi smearing of 0.05 eV is used for the geometry optimization and the further analysis. For SrCoO$_3$ and SrIrO$_3$, all atomic positions and unit cell parameters were optimized until the atomic forces were less than 0.01 eV/\AA.  For the strained DP Sr$_2$CoIrO$_6$ all atomic positions and the $c$-lattice constant were optimized.
  
  The optical spectrum of the strained double perovskite SCIO was obtained within the I.P. approximation. Additionally, quasiparticle effects were considered with a single shot $G_{0}W_{0}$ calculation\cite{Hedin}  and excitonic effects by solving the BSE\cite{Rohlfing, Albrecht}. For the calculation of the real and imaginary part of the dielectric function we used 50 frequency grid points  with a cut-off energy of 600 eV. A total of 48 valence and 54 conduction bands were employed for the calculation of the BSE spectrum, which ensures convergence of the spectrum up to at least 3 eV. A Gaussian broadening of 0.1 eV is used for the I.P. and $G_{0}W_{0}$ spectrum and 0.05 eV for the BSE spectrum. 
  
\subsection{\label{sec:ExpDetail}Experimental details}
The SCIO thin films investigated in this study were grown by the MAD technique on various pseudocubic (pc) (001) oriented substrates. MAD employs an oxygen-rich growth atmosphere and enables to prepare high-quality perovskite thin films \cite{Schneider, Egoavil, Jungbauer}. To protect the thin films from environmental degeneration, a STO capping layer was grown also by MAD directly after the SCIO film. The characterization of each sample, in particular, phase purity, degree of $B$-site ordering and degree of strain transfer, was done by room-temperature x-ray diffraction (XRD, XRR and RSM), far-field Raman spectroscopy and high-resolution scanning transmission electron microscopy: reciprocal space maps (RSMs) around the (013)$_\text{pc}$ substrate peak in combination with HAADF STEM images in [100]$_\text{pc}$ direction confirm the fully strained state of the SCIO thin films. The degree of B-site ordering was analyzed by XRD scans in tilted geometry as well as by polarization-dependent Raman spectroscopy. Together with HAADF STEM images in [110]$_\text{pc}$ direction of selected samples they indicate at least 65\% Co/Ir ordering. For more details on the characterization see Ref.~\cite{Gegenwart}.

The optical properties of SCIO, deposited on three different substrates [(LaAlO$_3$)$_{0.3}$(Sr$_2$TaAlO$_6$)$_{0.7}$ (LSAT), STO, and GSO], were investigated by optical transmission spectroscopy in vacuum, using a Bruker IFS 66v/S Fourier-transform infrared spectrometer. In order to cover the wide energy range of interest, we employed different detectors (silicon (Si), germanium (Ge), and mercury cadmium telluride (MCT)) and different beamsplitters (quartz and KBr). The spectra for the different energy regions were subsequently stitched together, using appropriate scaling factors, to obtain a single spectrum covering the energy range from 0.3 eV to 2.6 eV. To obtain the energy-dependent absorption coefficient, we used Lambert-Beer’s law: $\alpha = - ln(I/I_0)$, where $I$ is the transmission of the sample and $I_0$ is the transmission through a reference sample, i.e. a substrate without the SCIO layer. The resulting absorption spectra (see Fig.~\ref{fig:dielectric_function}), exhibit an almost linearly increasing background signal, leading to a slightly tilted spectrum. This background signal stems most likely from deviations between the reference and the SCIO sample including the STO capping layer and could not be reduced in these measurements.

\section{\label{sec:Results}Results And Discussion}
\subsection{\label{sec:strmag}Structural and magnetic properties}


\begin{figure}[!htp]
\includegraphics[width=0.47\textwidth]{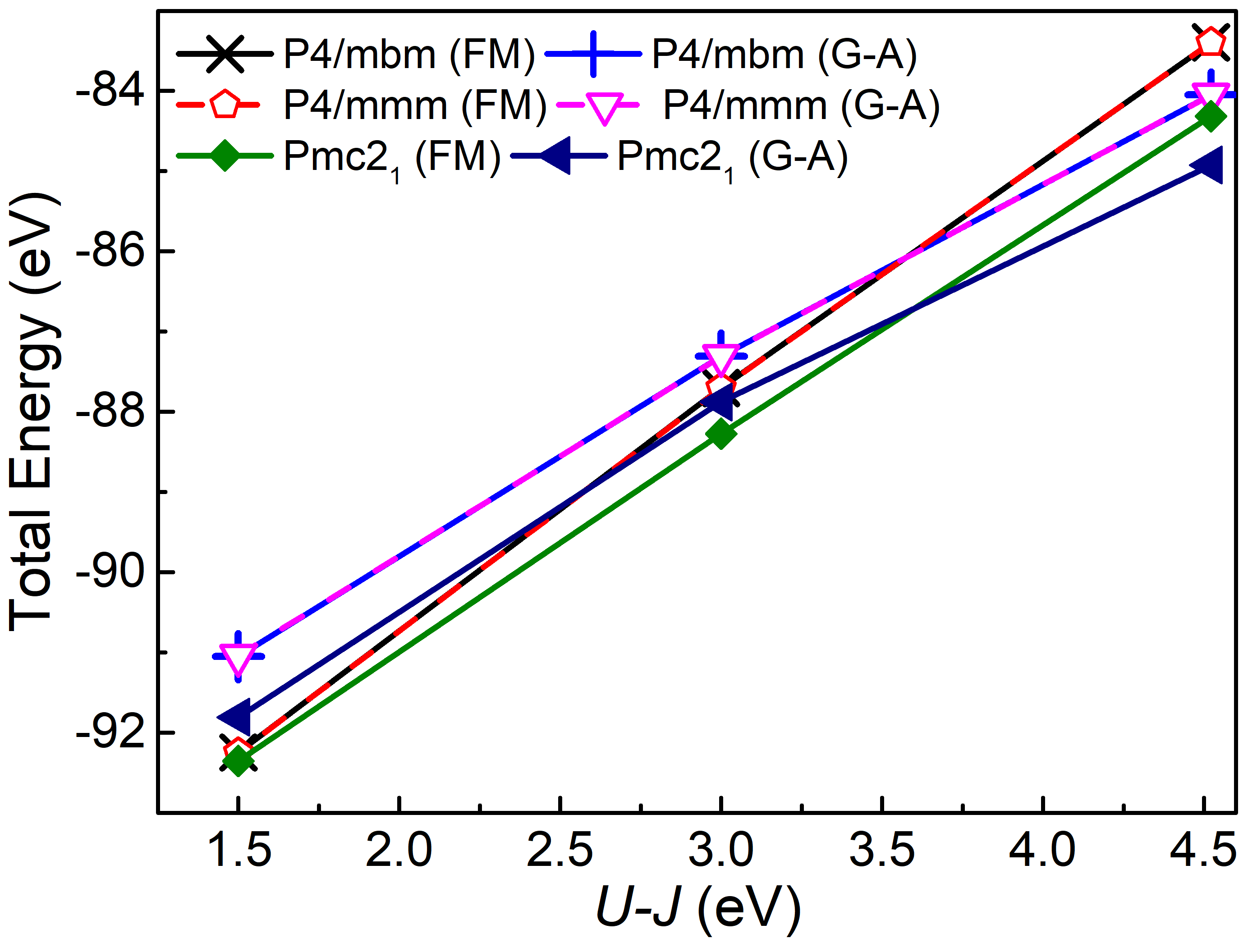}%
\caption{\label{fig:SCOenergy} The total energy of SrCoO$_3$ with FM and AFM order for different symmetry for $U$ = 1.5 eV, 3.0 eV and 4.52 eV.}
\end{figure}


\begin{figure*}[!htp]
\includegraphics[width=0.9\textwidth]{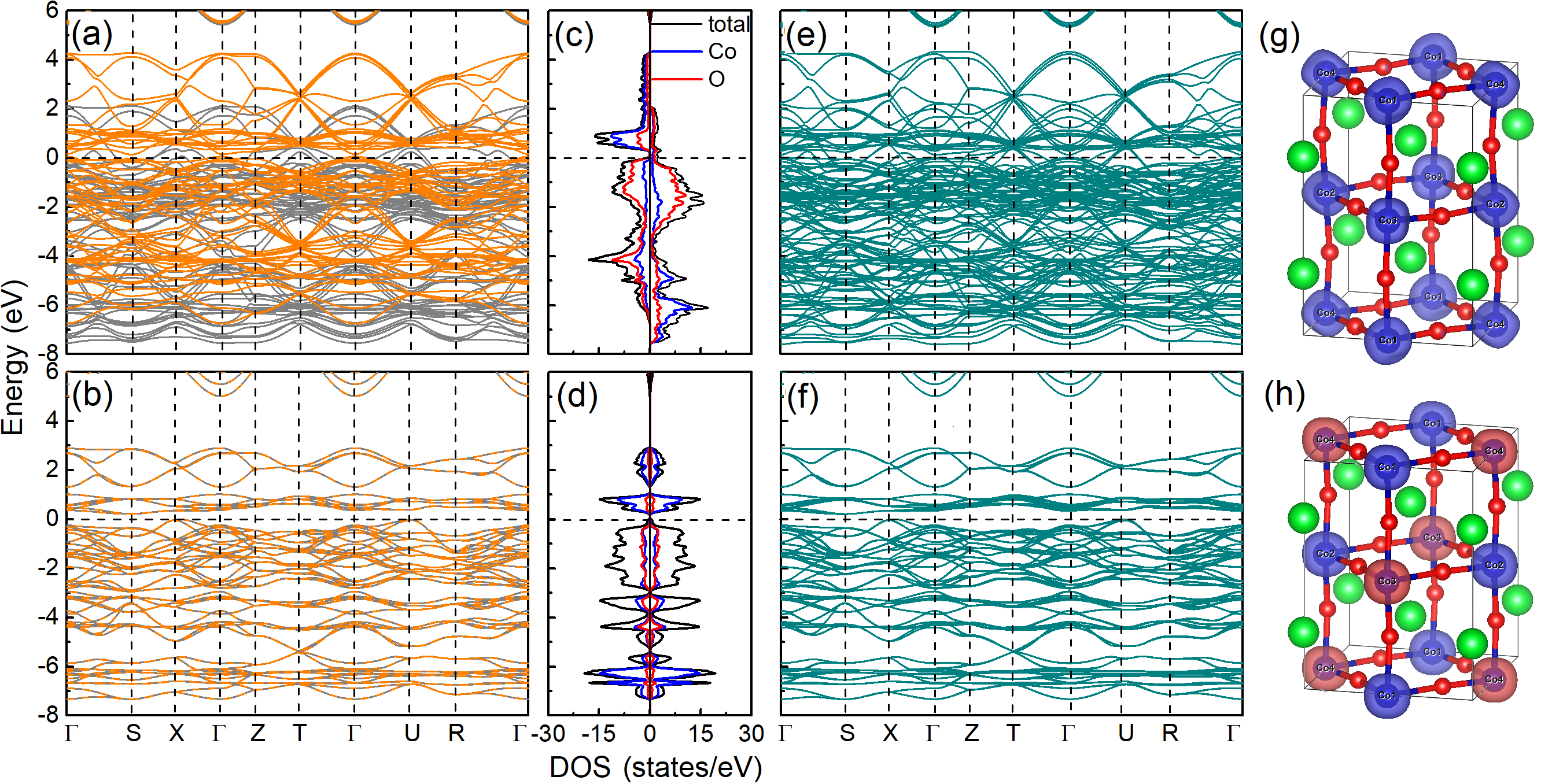}%
\caption{\label{fig:SCOband} DFT+$U$ band structure (a), (b) without and (c), (d) with SOC, element-projected density of states (PDOS) and (e), (f) spin density of SrCoO$_3$ for FM (top panels) and G-type AFM coupling (bottom panels). Majority and minority bands in (a) and (b) are shown in dark grey and orange, respectively.}
\end{figure*}

\begin{figure}[!htp]
\includegraphics[width=0.45\textwidth]{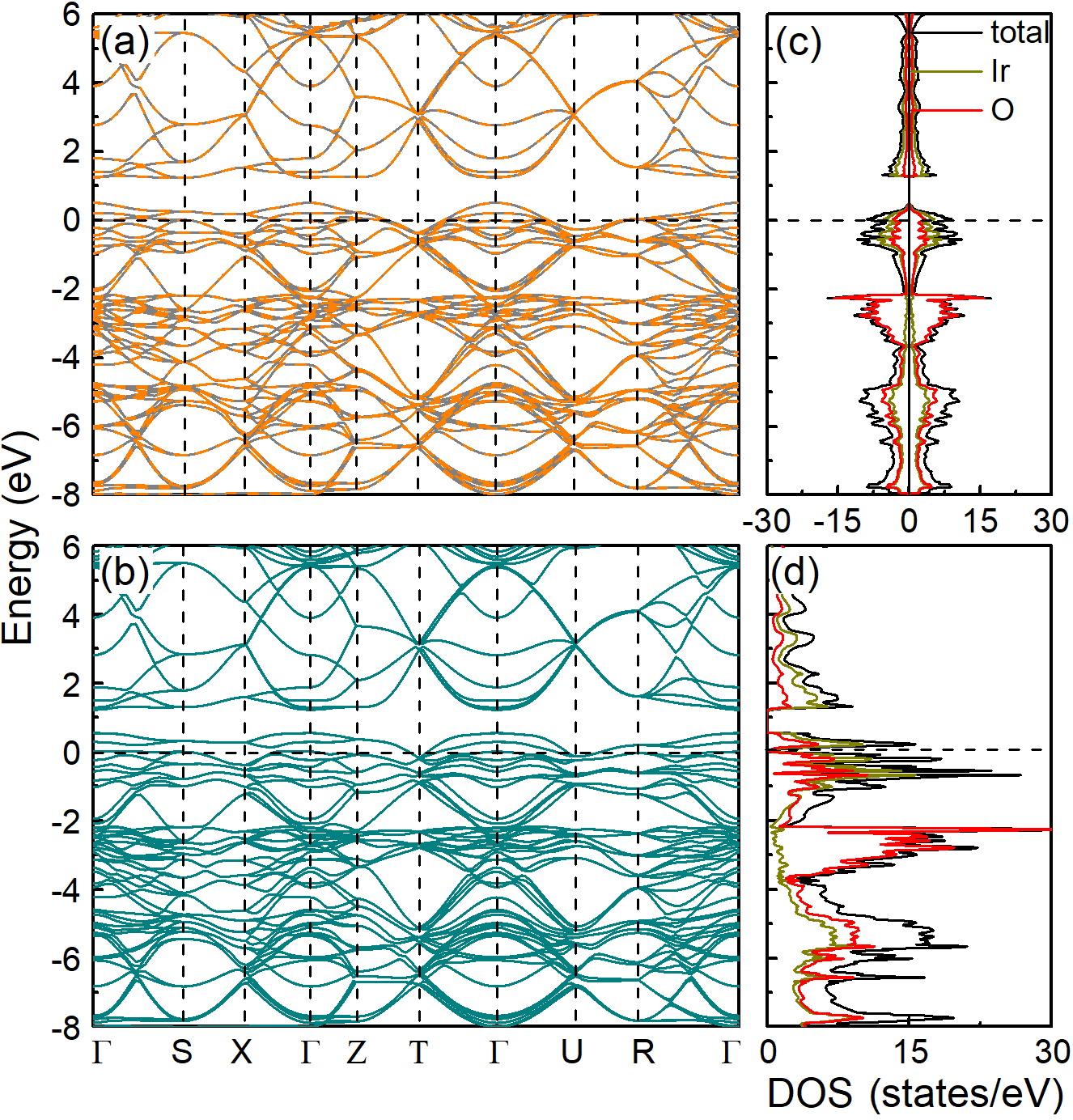}%
\caption{\label{fig:SIOband} DFT+$U$ band structure and element-projected density of states of SrIrO$_3$ (a) without and (b) with SOC. For the band structure the same color scheme was used as in Fig.~\ref{fig:SCOband}.}
\end{figure}

\begin{table*}[!htp]
\caption{\label{tab:withoutsoc} The DFT+$U$ values of lattice constants, band gaps, spin moments, out-of-plane and in-plane bond angle and bond length for SrCoO$_3$, SrIrO$_3$ and strained Sr$_2$CoIrO$_6$, without spin-orbit coupling (SOC). For comparison we list also the measured $c_{\rm exp}$ from Ref.\cite{Gegenwart}. }
\footnotesize
\begin{ruledtabular}
\begin{tabular}{cccccccccccccc}
{}&\multicolumn{4}{c}{Lattice constant (\AA)}&$E_g$(meV)&$M_S^{Co}$($\mu_{B}$)&$M_S^{Ir}$($\mu_B$)&\multicolumn{2}{c}{$M_1$-O-$M_2$($^{\circ}$)}&\multicolumn{4}{c}{$M$-O(\AA)}\\
\hline
{}&{}&{}&{}&{}&{}&{}&{}&{}&{}&\multicolumn{2}{c}{Co-O}&\multicolumn{2}{c}{Ir-O}\\
\hline
{}&$a$&$b$&$c$&$c_{\rm exp}$\cite{Gegenwart}&{}&{}&{}&$\|$&$\bot$&$\|$&$\bot$&$\|$&$\bot$\\
\hline
SrCoO$_3$(FM)&5.395&5.386&7.66&{}&0&2.51&{}&173&171&1.92/1.95&1.91&{}&{}\\
\hline
SrCoO$_3$(G-AFM)&5.528&5.517&7.48&{}&220&2.85&{}&171&168&1.80/2.12&1.89&{}&{}\\
\hline
SrIrO$_3$&5.54&5.64&7.85&{}&0&{}&0&154&154&{}&{}&2.03&2.02\\
\hline
SCIO$@a_{\rm NGO}$&7.73&7.73&7.99&8.05&163&2.76&1.31&158&165&2.00&2.06&1.94&1.96\\
\hline
SCIO$@a_{\rm STO}$&7.81&7.81&7.91&7.91&235&2.75&1.38&159&163&2.02&2.05&1.95&1.96\\
\hline
SCIO$@a_{\rm GSO}$&7.93&7.93&7.74&7.73&187&2.74&1.36&160&162&2.05&1.97&1.95&1.94\\
\end{tabular}
\end{ruledtabular} 
\end{table*}

\begin{table*}[!htp]
\caption{\label{tab:withsoc} The DFT+$U$ values of lattice constants, band gaps, spin and orbital moments for SrCoO$_3$, SrIrO$_3$ and strained Sr$_2$CoIrO$_6$, including spin-orbit coupling (SOC).}
\footnotesize
\begin{ruledtabular}
\begin{tabular}{ccccccccc}
{}&\multicolumn{3}{c}{Lattice constant (\AA)}&$E_g$(meV)&$M_S^{Co}$($\mu_{B}$)&$M_S^{Ir}$($\mu_B$)&$M_L^{Co}$($\mu_{B}$)&$M_L^{Ir}$($\mu_{B}$)\\
\hline
{}&$a$&$b$&$c$&{}&{}&{}&{}\\
\hline
SrCoO$_3$(FM)&5.414&5.407&7.61&0&2.50&{}&0.13&{}\\
\hline
SrCoO$_3$(G-AFM)&5.528&5.517&7.48&215&2.82&{}&0.09&{}\\
\hline
SrIrO$_3$&5.56&5.62&7.88&0&{}&0&{}&0\\
\hline
SCIO$@a_{\rm NGO}$&7.73&7.73&7.90&144&2.45&1.21&0.31&0.13\\
\hline
SCIO$@a_{\rm STO}$&7.81&7.81&7.87&189&2.35&1.22&0.46&0.13\\
\hline
SCIO$@a_{\rm GSO}$&7.93&7.93&7.78&151&2.40&1.25&0.32&0.14\\
\end{tabular}
\end{ruledtabular} 
\end{table*}

The structural, electronic and magnetic properties of the three compounds obtained from the DFT+$U$ calculations without and with SOC are summarized in Table~\ref{tab:withoutsoc} and \ref{tab:withsoc}, respectively. We begin the discussion with the end members SrCoO$_3$ and SrIrO$_3$, before turning to the double perovskite Sr$_2$CoIrO$_6$.  

For SCO we considered the tetragonal phases ($P4/mbm$ and $P4/mmm$) proposed in a previous PBE+$U$ study~\cite{Cazorla}, however, full structural optimization rendered an orthorhombic $Pmc2_1$ phase, 0.57 and 0.56 eV per simulation cell more stable than the tetragonal ones. Furthermore, the magnetic order of the ground state is sensitive to the Hubbard $U$ parameter: The energetic stability of the different phases, plotted in Fig.~\ref{fig:SCOenergy} as a function of $U$, shows that  ferromagnetic coupling is favored for $U\leq 3$eV, but for higher values it switches to a G-type antiferromagnetic state, with significant B-site off-centering, likely related to the AFM ferroelectric  phase predicted by Lee and Rabe~\cite{Rabe}. Since the ground state of SCO is FM metallic~\cite{Tokura}, we proceed the further analysis using $U=3$ eV for Co. 
The band structure, projected density of states (PDOS) and spin density for the FM and AFM phases  is displayed in Fig.~\ref{fig:SCOband}.  
For $U=3$~eV, the FM metallic state is preferred by 0.40 eV compared to the AFM-G insulating phase, with Co in an intermediate spin state ($t_{2g}^4 e_g^1$) with a magnetic moment of 2.51 and 2.85 \mub\ for the FM and AFM phases, respectively. Including spin-orbit coupling, Co acquires an orbital moment of 0.13~\mub\ (FM) and 0.09~\mub\ (AFM). Furthermore, the [110]-magnetization direction is favored by 8.34 meV over [001]. Concerning the structural properties, SCO shows noticeable octahedral tilts (cf. Fig.~\ref{fig:structure} a) expressed in  a reduction of the in-plane and out-of-plane Co-O-Co angles to 173$^{\circ}$ and 171$^{\circ}$, respectively, whereas the in- and out-of-plane Co-O bond lengths are 1.92 and 1.90~\AA, respectively. For the AFM-G phase three distinct bond lengths are obtained: 1.80 and 2.12~\AA) in-plane and 1.89~\AA\ out-of-plane.
 
The other end member, SIO, exhibits a $Pbnm$ structure with an ($a^-a^-c^+$) tilt pattern in Glazer's notation~\cite{Rondinelli} and a strongly reduced Ir-O-Ir bond angle of 154$^{\circ}$. The lattice parameters ($a=5.54$~\AA, $b=5,64$~\AA\ and $c=7.85$~\AA) are in good agreement with previous studies~\cite{Rondinelli} with an Ir-O bond length of 2.02~\AA. Both the spin and orbital moment of Ir$^{4+}$ ($d^5$) are quenched and the system is semimetallic with a fully occupied $j=3/2$ and half-filled  $j=1/2$ band \cite{Kee, Franchini}.

To model the growth of the double perovskite Sr$_2$CoIrO$_6$ on different substrates used in experiment~\cite{Gegenwart} we have strained the lateral lattice constant to the pseudocubic one of NdGaO$_3$ (NGO), $a_{\rm NGO}=3.865$ \AA, SrTiO$_3$ (STO), $a_{\rm STO} = 3.905$ \AA, and GdScO$_3$ (GSO), $a_{\rm GSO}=3.965$~\AA, which corresponds to compressive strain, unstrained case and tensile strain, respectively. As expected, with increasing lateral lattice constant the optimized $c$-parameter decreases from 7.99~\AA\ ($a_{\rm NGO}$) to 7.91~\AA\ ($a_{\rm STO}$) and, finally, 7.74~\AA\ ($a_{\rm GSO}$), in close agreement with the measured values~\cite{Gegenwart}. The  bond angles exhibit values between those of SCO and the strongly distorted SIO: while the in-plane Co-O-Ir angle increases with $a$ from 158$^{\circ}$ to 160$^{\circ}$, the out-of-plane angle decreases from 165$^{\circ}$ to 162$^{\circ}$. Concerning the bond lengths, a considerable enhancement compared to the SCO end member is observed for Co-O with a noticeable difference of the in- and out-of-plane distances  at $a_{\rm NGO}$ (2.00, 2.05~\AA) which is nearly quenched for $a_{\rm STO}$ (2.02, 2.05~\AA) and  reversed for $a_{\rm GSO}$ (2.07, 1.96~\AA). In contrast, the Ir-O bond lengths are significantly reduced and show a weaker dependence on strain with in- and out-of-plane values of 1.94 and 1.96~\AA\ for compressive strain, 1.95 and 1.96~\AA\  at  $a_{\rm STO}$, and 1.95 and 1.94~\AA\ for $a_{\rm GSO}$. These notable changes of bond lengths in SCIO w.r.t the end members are closely related to the electronic reconstruction and charge transfer, that will be discussed in subsection~\ref{sec:elprop}.  

\begin{figure*}[!htp]
\includegraphics[width=0.9\textwidth]{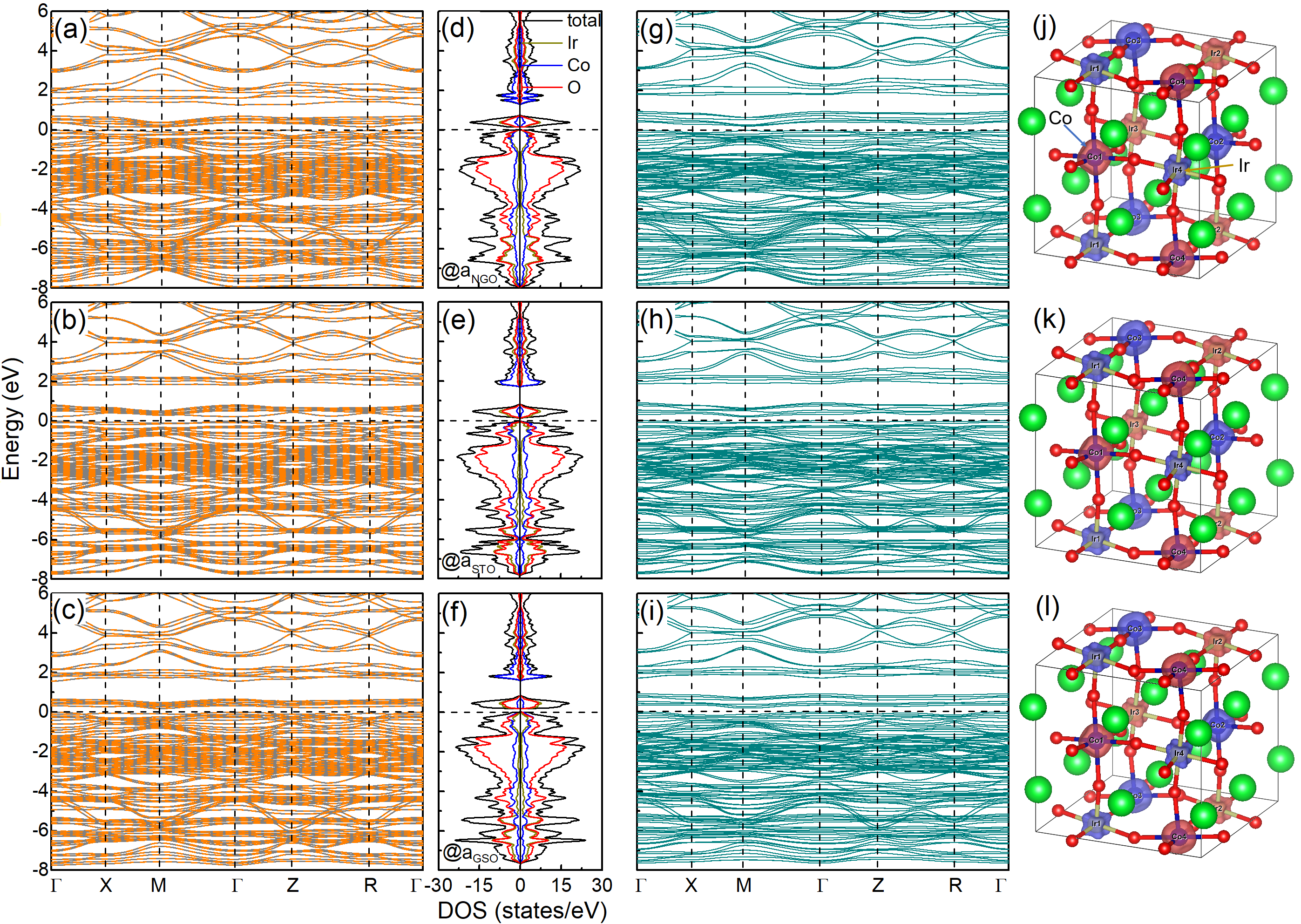}%
\caption{\label{fig:SCIOband} DFT+$U$ band structure, projected density of states (PDOS) and spin density of Sr$_2$CoIrO$_6$ (a)-(c) without and (d)-(f) with SOC strained at: (a) and (d) $a_{\rm NGO}$, (b) and (e) $a_{\rm STO}$, (c) and (f) $a_{\rm GSO}$. For the band structure the same color scheme was used as in Fig.~\ref{fig:SCOband}.}
\end{figure*}
 
\begin{figure*}[!hbt]
\includegraphics[width=0.95\textwidth]{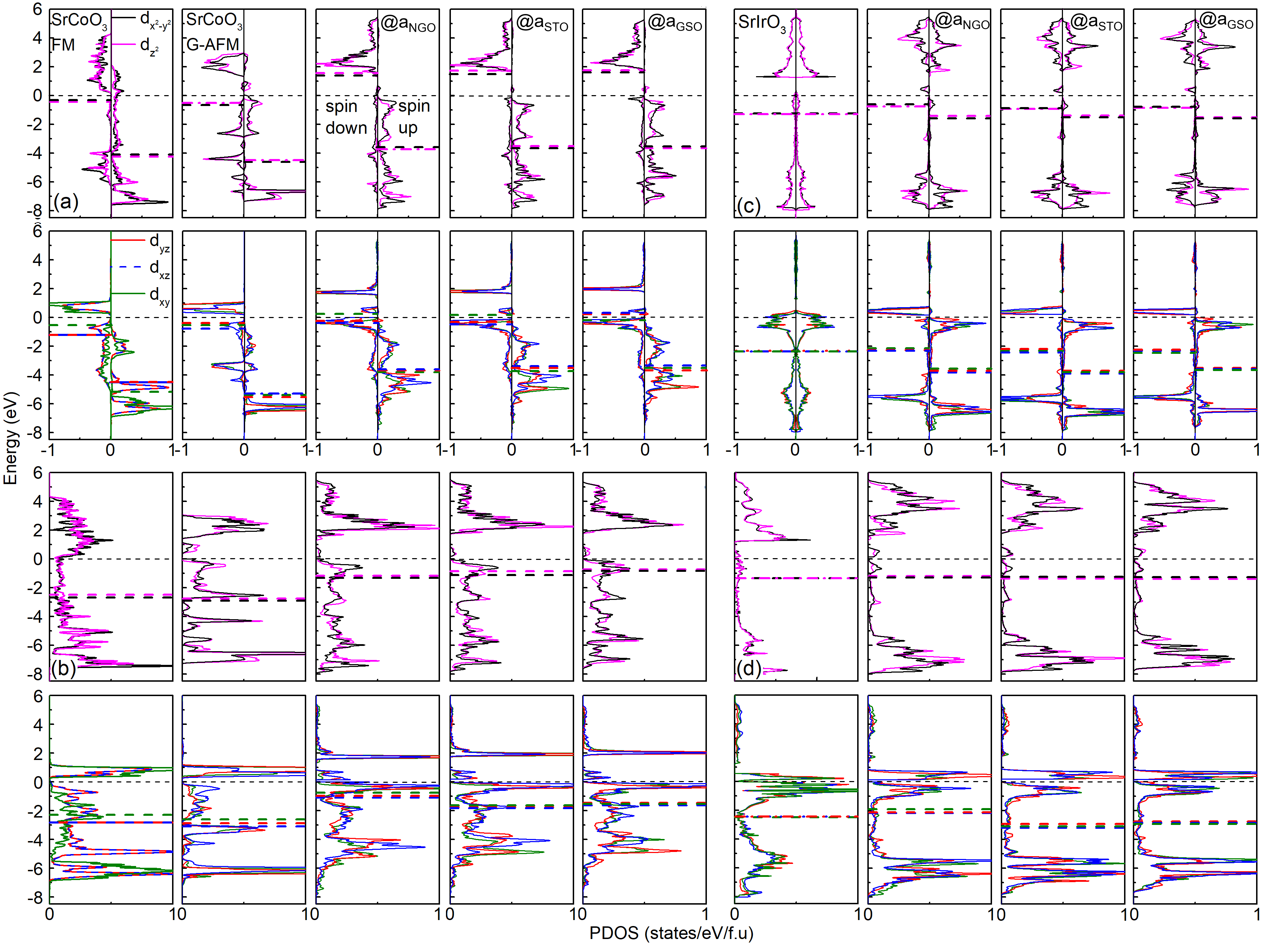}%
\caption{\label{fig:pdos} Orbitally projected DOS of Co $3d$ states in SrCoO$_3$ and strained Sr$_2$CoIrO$_6$ (a) without  and (b) with SOC, and Ir $5d$ states in SrIrO$_3$ and Sr$_2$CoIrO$_6$ (c) without  and (d) with SOC. For better visibility,  the \eg\ and \tg\ states are shown in separate panels. Additionally, the band-center of each $d$ orbital is given by a dashed line.}
\end{figure*}

SCIO is an antiferromagnetic insulator, 0.50-0.57 eV per simulation cell more favorable than the ferromagnetic phase. The AFM configuration comprises mixed Co, Ir planes that couple antiparallel along the  [110]-direction, as can be seen from the spin density plotted in Fig.~\ref{fig:SCIOband}(g)-(i). This configuration was found to be more stable than AFM order along the [111] or [001]-directions in preliminary calculations\cite{Moss2017} with the WIEN2k code~\cite{wien2k}, and is consistent with the magnetic order determined from neutron powder diffraction.~\cite{EhrenbergH}. The spin moment of Co in SCIO  ($\sim2.74$~\mub) is comparable to the one in bulk SCO [2.51~\mub\ (FM) and 2.85~\mub\ (AFM)] and close to the experimental (2.6(1)~\mub) and theoretical value ($\sim3.00$~\mub) reported in Ref.~\cite{EhrenbergH}. Most interestingly, Ir acquires a significant magnetic moment of $1.31-1.38$~\mub, compared to the zero magnetic moment in bulk SIO and at variance with a vanishing contribution of Ir suggested from experiment~\cite{EhrenbergH} (the theoretical value from that study is $\sim0.81$~\mub). The origin of this effect will be discussed in the next subsection.  Including SOC, the spin moments of both ions are somewhat reduced and show a weak dependence on strain: 2.35-2.45~\mub\ for Co and 1.21-1.25~\mub\ for Ir with increasing lateral lattice constant. At the same time the orbital moments are substantially enhanced w.r.t. the end members and vary between 0.31-0.46~\mub\ for Co (the latter at $a_{\rm STO}$) and 0.14-0.13~\mub\ for Ir. Recent XAS measurements~\cite{Gegenwart} suggest a switching of the magnetic easy axis. While our results do not show a reversal, the energy difference between the preferred [100]-magnetization direction  over the [001] decreases from 85 meV to 70 meV per 40-atom u.c. from compressive to tensile strain.

\subsection{\label{sec:elprop}Electronic properties}

We now turn to the electronic properties. Figs.~\ref{fig:SCOband}, \ref{fig:SIOband} and \ref{fig:SCIOband} display the band structure, total and element-projected DOS with and without SOC for the end members SCO, SIO and the strained double perovskite SCIO, respectively. As mentioned above, SrCoO$_3$  is a ferromagnetic metal (Fig.~\ref{fig:SCOband} a and c), whereas the G-type AFM phase (Fig.~\ref{fig:SCOband} b and d) has an indirect band gap of 220 meV  with the top of the valence band at X and the bottom of the conduction band at S. 
SrIrO$_3$  is a metal that turns into a semimetal with a dip of the DOS at the Fermi level upon taking into account SOC (Fig.~\ref{fig:SIOband}). On the other hand, the double perovskite SCIO (Fig.~\ref{fig:SCIOband}) is an AFM insulator with a direct band gap at $\Gamma$ that shows a nonmonotonic behavior with strain and changes from 163 meV at $a_{\rm NGO}$ to 235 meV at $a_{\rm STO}$ and 187meV at $a_{\rm GSO}$. The band structure is weakly affected by SOC, the band gaps are somewhat reduced: 144 meV, 189 meV and 151 meV for $a_{\rm NGO}$, $a_{\rm STO}$ and $a_{\rm GSO}$, respectively. The influence of many body effects on the band gap will be discussed in Section~\ref{sec:optical}.

To gain further insight into the changes in orbital occupation, we  have plotted in Fig.~\ref{fig:pdos} the  orbitally projected DOS of the Co 3$d$  and Ir 5$d$ states both with and without SOC in the end members SCO, SIO and the double perovskite SCIO. For SCO both the FM metallic and AFM insulating phases are shown, but the latter allows to track more straightforward changes from SCO to  SCIO and is therefore used in the further analysis. For Co in SCO all $3d$ orbitals  are occupied in one spin channel, while there is a partial occupation in the other. Moreover, the two \eg\ and the three \tg\ orbitals are nearly degenerate. In contrast, in the strained DP SCIO this degeneracy is lifted. In particular the band width of the \eg\ states is enhanced for compressive strain, even stronger for the \dxtyt-orbital, that extends to lower energies at $a_{\rm NGO}$, whereas at  $a_{\rm GSO}$  the bottom of the  \dzt-band lies lower. 
The position and occupation of Co $3d$-bands changes significantly from SCO to SCIO: while the narrow unoccupied part of the \tg\ bands in the majority spin-channel of SCO lies between 0.2 and 1.1 eV and the broader \eg\ bands between 1-3 eV, in SCIO the former splits into an occupied part just below \ef\ and an unoccupied fraction  shifted to $\sim 1.5$ eV. Likewise, the empty \eg\ bands broaden and move upwards, ranging between 1.5 and 5.4 eV. The orbitally- and spin-resolved d-band centers, calculated according to Ref.~\cite{dband} are also displayed in Fig.~\ref{fig:pdos} and show an upward shift  both for Co and Ir compared to the end members. The d-band centers allow also a qualitative comparison to the level diagram resulting from the x-ray absorption spectra in Ref.~\cite{Gegenwart}: specifically, the band center of the \dxy\ orbital lies higher for compressive strain but a downward shift and reversal of sequence takes place for tensile strain, which is consistent with Ref.~\cite{Gegenwart}. 

A strong electronic rearrangement is observed for Ir in SCIO compared to the end member SIO, as shown in Fig.~\ref{fig:pdos}c: In SIO the PDOS around \ef\ is dominated by the partially occupied and degenerate \tg\ states in both spin channels, whereas the  unoccupied \eg\ bands extend between 1.3 and 5.3 eV. In contrast, in SCIO the latter move upwards and are narrower, lying between 1.6 and 5 eV. More severe changes occur for the \tg\ states: in one spin channel the unoccupied part splits off and shifts upwards localized between 0.17 and 0.88 eV, whereas in the other the \tg\ bands are fully occupied. The substantial spin splitting (also for the \eg\ bands) and imbalance in occupation of the \tg\ states explains the emergence of a significant spin moment at the Ir-sites of $\sim 1.3$~\mub, as compared to the nonmagnetic SIO. The effect of strain on the occupied Ir \tg\ bands is similar to the one on Co: the \dxy\ and \dxz\ orbitals prevail at the bottom of the \tg\ band for compressive and tensile strain, respectively,  whereas the \dzt\ states get narrower for tensile strain.  Similar effects are observed when including SOC, the major difference being the emergence of semimetallic behavior for bulk SIO when SOC is switched on.

\subsection{\label{sec:elpropCT}Quantification of charge transfer}
\begin{figure}[!htp]
\includegraphics[width=0.47\textwidth]{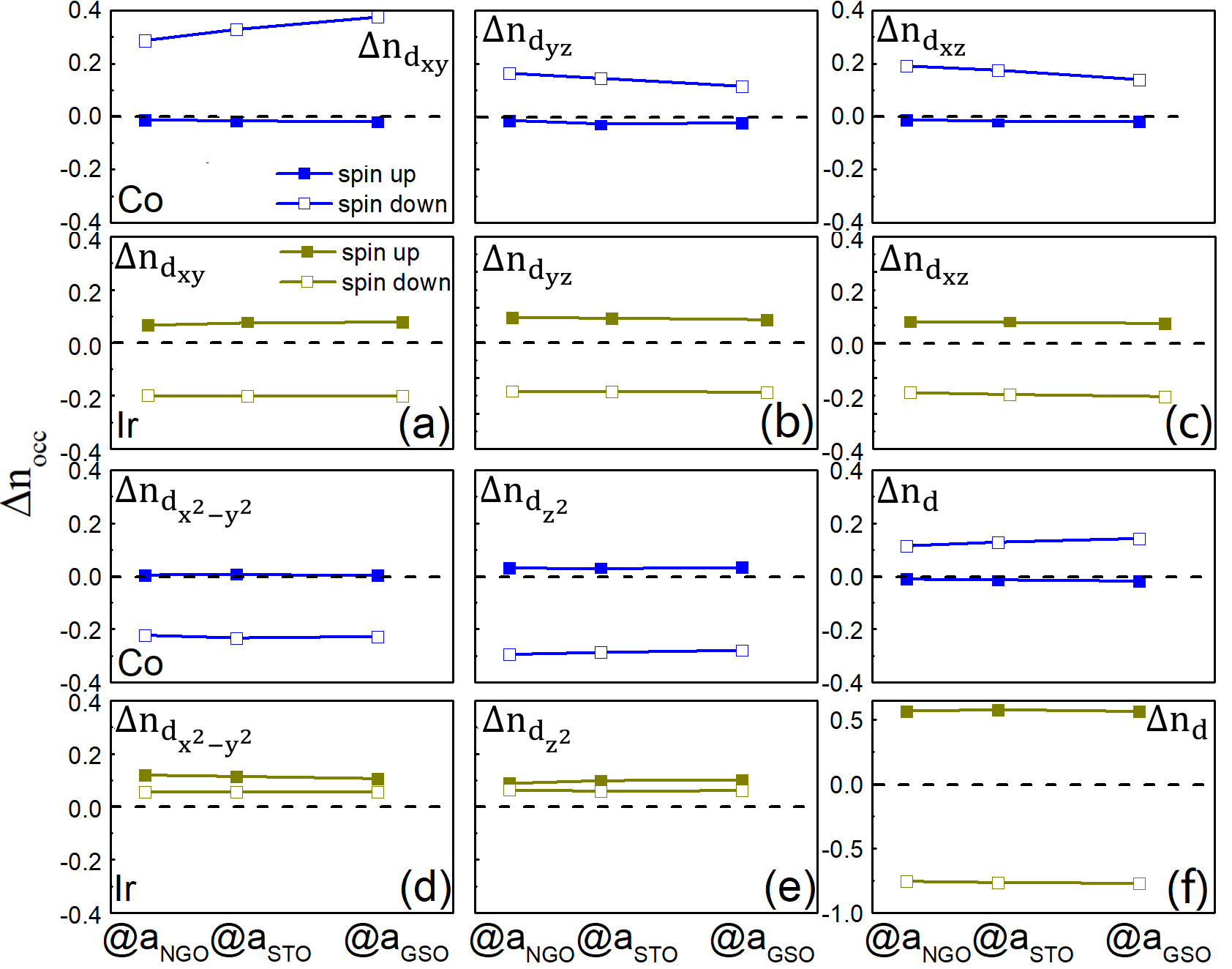}%
\caption{\label{fig:occupationwithoutsoc} Changes in occupation number of the five Co $3d$ (blue) and Ir $5d$ (yellow)  orbitals in SCIO with respect to the occupation number in the end members SCO, SIO without SOC as a function of strain. Solid and empty symbols represent changes in occupation numbers for spin up and spin down states, respectively.}
\end{figure}

\begin{figure}[!htp]
\includegraphics[width=0.47\textwidth]{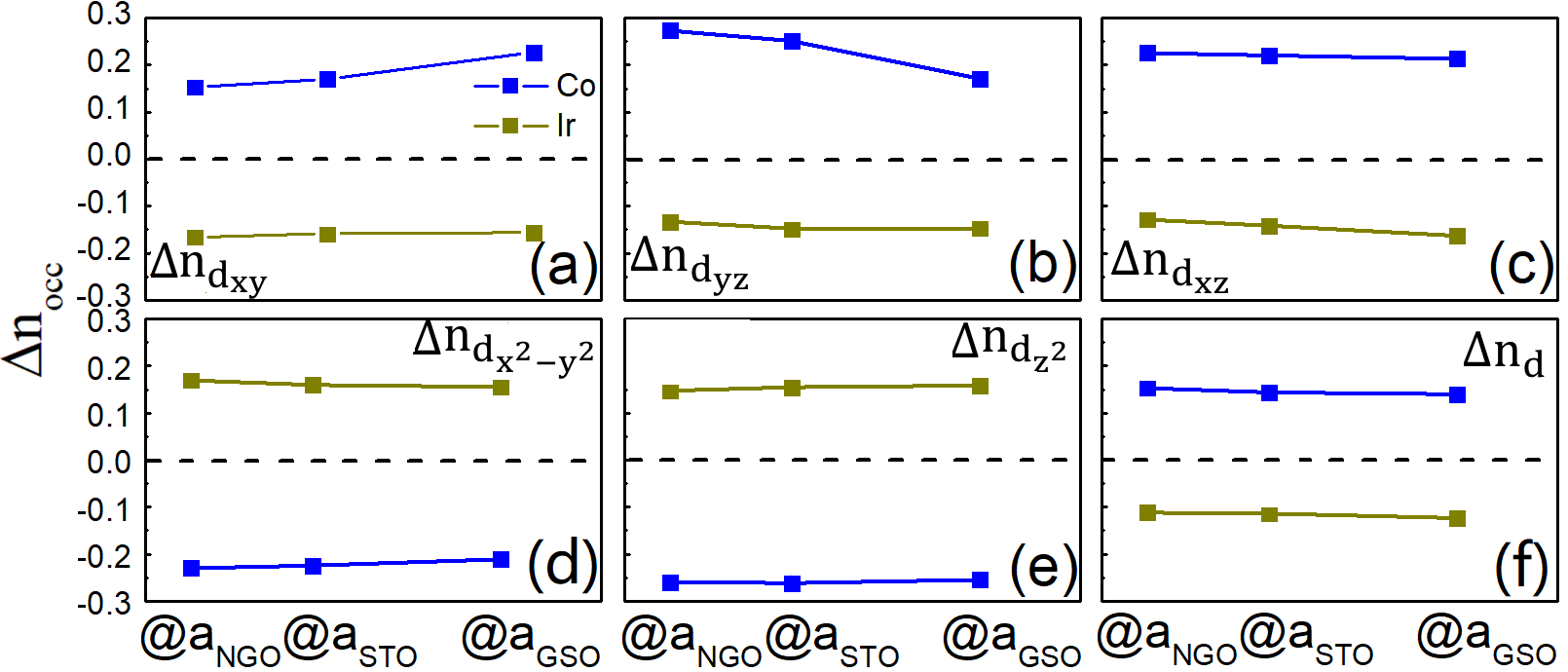}%
\caption{\label{fig:occupationwithsoc} Changes in occupation number of the five  Co $3d$ (blue) and Ir $5d$ (yellow) in SCIO with respect to the occupation number in the end members SCO, SIO with SOC as a function of strain.}
\end{figure}

Both the emergence of a substantial moment of $\sim1.3$~\mub\ at the Ir sites and the analysis of the orbitally-projected DOS above signal strong rearrangement in occupation and charge transfer at the Co and Ir sites in SCIO in comparison to the end members. To quantify the charge transfer we have plotted in Figs.~\ref{fig:occupationwithoutsoc} and ~\ref{fig:occupationwithsoc} the change in Co $3d$ and Ir $5d$ orbital occupation in SCIO w.r.t. the end members without and with spin-orbit coupling. The main effect is a strong reduction of  occupation by $\sim0.75e^-$ in the Ir \tg\ minority spin channel which is transferred partially ($\sim0.58e^-$) to the majority spin orbitals of Ir (mainly \tg, to a lesser extent \eg) and to Co ($\sim 0.15 e^-$). The latter effect is weaker than the previously proposed valence change from 4+ in the bulk end members to a Co$^{3+}$ and Ir$^{5+}$ configuration~\cite{Ehrenberg,Gegenwart,Tjeng} in SCIO. In contrast to the Ir $5d$ orbitals, the Co $3d$ occupation reveals substantial orbital  dependence in the minority spin channel (the majority being filled): both  \eg\ orbitals exhibit a reduction by 0.22$e^-$ (\dxtyt) and 0.30$e^-$ (\dzt), respectively. Simultaneously, the minority-spin \tg\ states occupation is enhanced, showing also a pronounced  strain-dependence:  while the \dxy-occupation increases by 0.28 to 0.38$e^-$ from compressive to tensile strain, the other two increase by 0.19 to 0.14$e^-$ (\dxz) by 0.0.15 to 0.11 (\dyz). Remarkably, these orbitally resolved changes with strain largely compensate in the total $d$ occupation, which shows no notable strain dependence. 

The changes in orbital occupation with SOC (Fig.~\ref{fig:occupationwithsoc}) are consistent with  the trends observed for the case without SOC and highlight the opposite tendency for Ir and Co: for the \tg\ orbitals there is a gain in occupation for Co (0.15-0.23$e^-$ for \dxy, 0.22-0.21$e^-$ for \dxz, and 0.27-0.17$e^-$ for \dyz) and loss for Ir (-0.15 for \dxy, -0.14$e^-$ for \dxz\ and \dyz), this trend is reversed for the \eg\ states: Ir shows a weak enhancement of 0.15$e^-$ each, whereas Co a reduction in electron count ($\sim 0.24 e^-$ each). The total $d$ occupation indicates a net charge transfer of 0.15$e^-$ from Ir to Co.

\begin{figure}[!htp]
\includegraphics[width=0.47\textwidth]{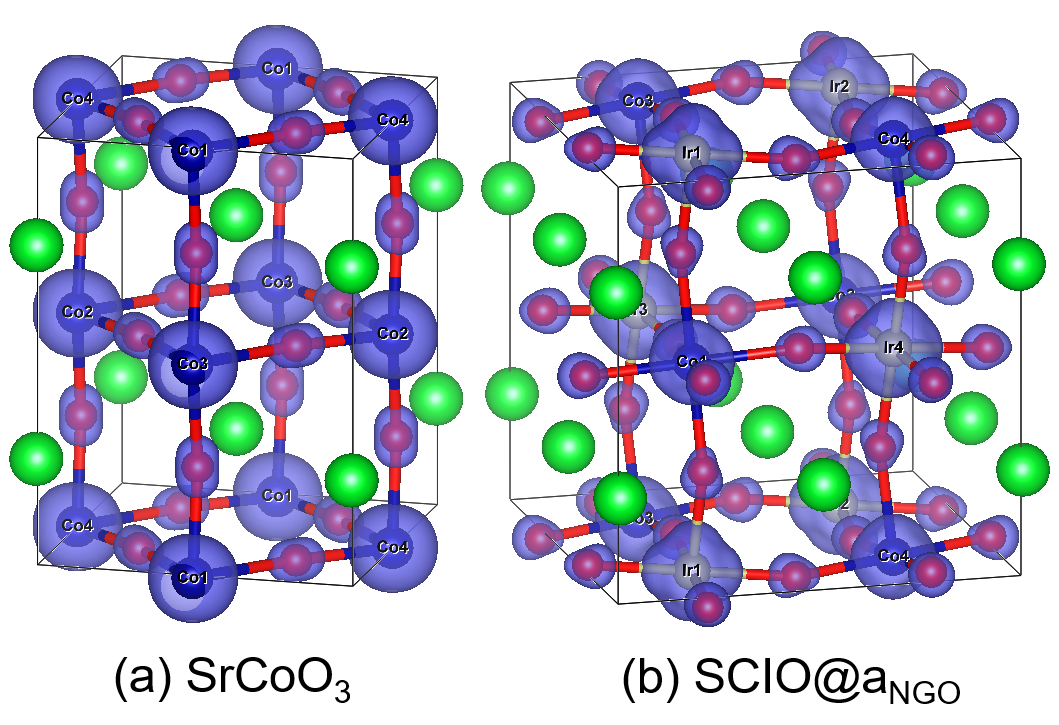}%
\caption{\label{fig:hole density} Spatial distribution of the unoccupied states integrated between the Fermi level and 6.0 eV without SOC (Isosurface value is 0.06$e^-$ \AA$^{-3}$) for (a) the end member SrCoO$_3$ for G-AFM coupling and (b) the double perovskites (b) SCIO at $a_{\rm NGO}$ as a representative example.}
\end{figure}


To visualize the electronic reconstruction as well as the involvement of the ligands, we have plotted in Fig.~\ref{fig:hole density}  the spatial distribution of the unoccupied states in SCO and SCIO, integrated from the Fermi energy to 6.0 eV.  In SCO, a significant nearly spherical hole density is observed around the Co sites as well as a substantial contribution at the ligand sites, signifying of the covalency of the bond. In contrast in SCIO, the density distribution of the unoccupied states around Co is more anisotropic and somewhat reduced, indicative of the charge transfer towards Co. Moreover, the electron density at the ligand sites pointing towards Co is nearly quenched,  whereas the lobes pointing towards the Ir site are present. This polarization at the ligands in SCIO underpins the enhanced ionicity of the Co-O bond, whereas the covalent contribution of the Ir-O bond persists. 


\begin{figure*}[!htb]
\includegraphics[width=0.9\textwidth]{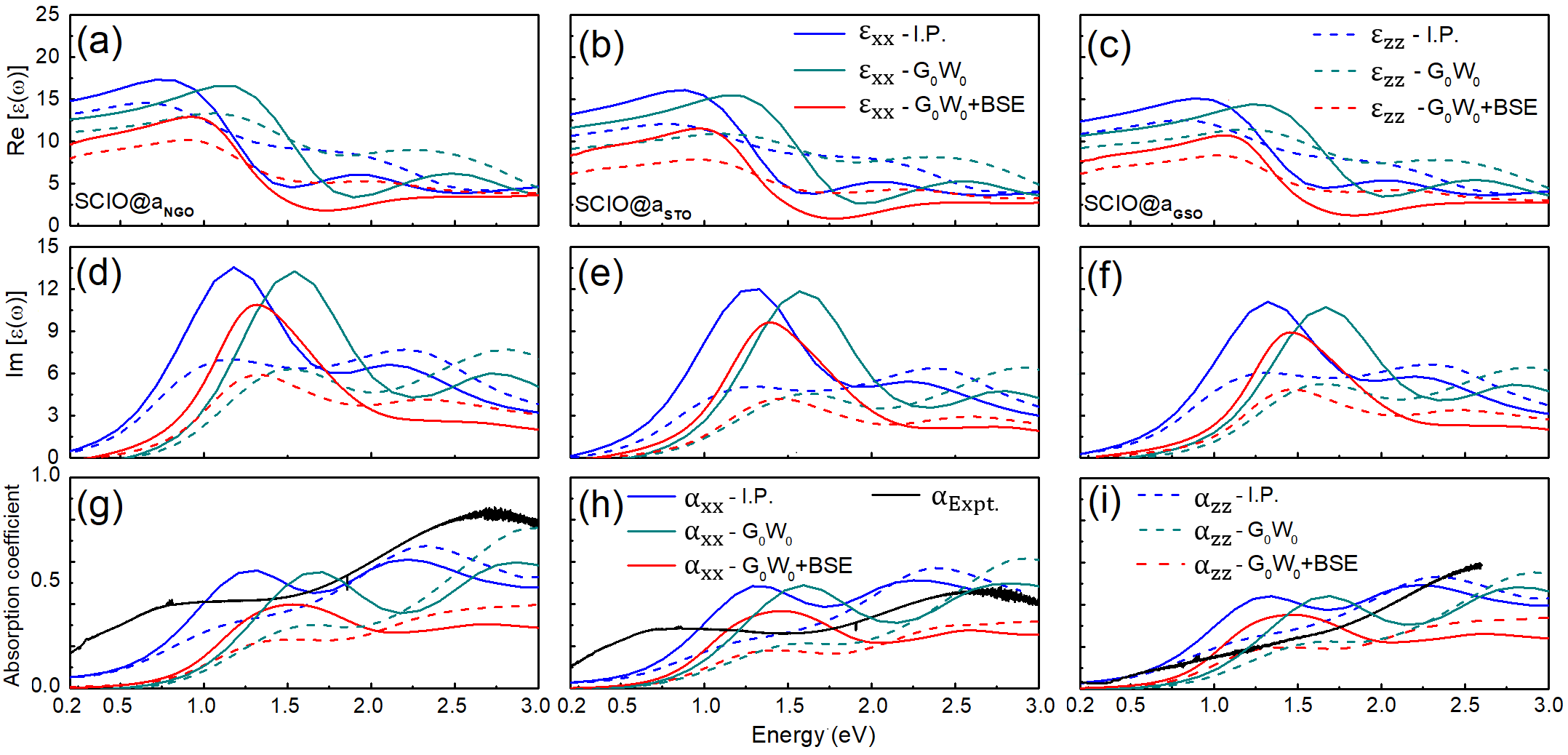}%
\caption{\label{fig:dielectric_function} (a)-(c) Real and (d)-(f) imaginary part  of the dielectric tensor within the I.P. (blue line), $G_{0}W_{0}$ (petrol line) and $G_{0}W_{0}$+BSE (red line) approximations for SCIO  strained at (a), (d) $a_{\rm NGO}$, (b), (e) $a_{\rm STO}$ and (c), (f) $a_{\rm GSO}$. The  in- and out- of plane components are denoted by solid and dashed lines, respectively. (g)-(i) The absorption coefficient in comparison with the one obtained from transmission measurements (black line). We note that in experiment for compressive strain a sample grown on an LSAT substrate was used instead of NGO, however both have similar (pseudocubic) lattice constant: $a_{\rm LSAT}=3.868$~\AA\ vs. $a_{\rm NGO}=3.855$~\AA, respectively~\cite{Gegenwart}.
}
\end{figure*}

\subsection{\label{sec:optical}Optical properties}

  The optical properties of the strained double perovskite SCIO  are assessed  in the I.P. approximation, as well as including quasiparticle and excitonic effects  by performing a $G_0W_0$ and BSE calculation for the case without SOC. The real and imaginary parts of the dielectric tensor are displayed in Fig.~\ref{fig:dielectric_function}.  Additionally, in the lower panels the absorption coefficient, derived from Re[$ \varepsilon$($\omega$)] and  Im[$ \varepsilon$($\omega$)],   is compared  to the one obtained from transmission measurements.

To explore the effect of strain, the in- and out-of-plane components of Re[$ \varepsilon$($\omega$)] and  Im[$ \varepsilon$($\omega$)] are plotted with solid and dashed lines, respectively, in Fig.~\ref{fig:dielectric_function}. The spectrum obtained within the I.P. approximation, exhibits a two-peak structure with the first peak at $\sim 1.2$~eV and the second at $\sim 2.0-2.4$~eV. Recalling the band structure in Fig.~\ref{fig:SCIOband} and the orbitally projected DOS in Fig.~\ref{fig:pdos}, we conclude  that the first peak corresponds to transitions between the top of the valence band and the narrow unoccupied \tg\ states that extend between 0.16-1.0 eV, whereas the second peak originates from transitions to the broader \eg\ states above 1.7 eV. At  $a_{\rm NGO}$ the first peak in Im[$ \varepsilon_{xx}$($\omega$)]  is predominant, more than two times higher than the second and than the peaks of Im[$ \varepsilon_{zz}$($\omega$)]. In the latter the the second peak is slightly more pronounced than the first and  is also shifted to higher energy. The main effect of strain is the decrease of the height of the first peak from compressive to tensile strain, thus decreasing the peak height difference in Im[$ \varepsilon_{xx}$($\omega$)] and also to the peaks in Im[$ \varepsilon_{zz}$($\omega$)]. 

Inclusion of quasiparticle effects within the $G_{0}W_{0}$ approximation shifts the spectrum to higher energies by $\sim 0.34$ eV, and lowers the height of the second peak in Im[$ \varepsilon_{xx}$($\omega$)], the opposite effect occurs for Im[$ \varepsilon_{zz}$($\omega$)], where the second peak prevails.  Excitoning effects, considered by solving the Bethe-Salpeter equation lead to a redshift of the spectrum. Additionally, the height of the first peak in Im[$ \varepsilon_{xx}$($\omega$)] is reduced and the second peak  is suppressed, yielding  only one main peak, followed by a broader plateau. On the other hand, Im[$ \varepsilon_{zz}$($\omega$)] exhibits a slightly more pronounced first peak, whose height is $\sim 50$\% of the one of Im[$ \varepsilon_{xx}$($\omega$)] and decreases for $a_{\rm STO}$ and  subsequently increases for $a_{\rm GSO}$.

As discussed previously, the band gap in the I.P. case shows a nonmonotonic behavior as a function of strain: 0.163 eV ($a_{\rm NGO}$), 0.235 eV ($a_{\rm STO}$) and 0.187 eV ($a_{\rm GSO}$). Including quasiparticle effects leads to a blueshift of the spectrum to higher energies, increasing the band gap to 0.503 eV ($a_{\rm NGO}$), 0.576 eV ($a_{\rm STO}$) and 0.529 eV ($a_{\rm GSO}$). Upon taking into account excitonic effects by solving the BSE, spectral weight is redistributed to lower energies and the onset of the spectrum lies between the ones for I.P. and $G_0W_0$ at 0.352 eV ($a_{\rm NGO}$), 0.458 eV ($a_{\rm STO}$) and 0.386 eV ($a_{\rm GSO}$).

To enable a comparison with experiment, in Fig.~\ref{fig:dielectric_function} (g)-(i) we plot  the absorption coefficient. Within I.P. the in-plane absorption coefficient shows a two peak structure with a comparable size of the peaks, whereas the out-of-plane coefficient has rather a two-plateau shape. The relative height difference between the in- and out-of-plane coefficient is reduced as the biaxial strain changes from compressive to tensile. While $G_{0}W_{0}$ leads mainly to a blueshift of the spectrum, subsequent inclusion of excitonic effects redshifts slightly the spectrum and suppresses nearly the second peak in $\alpha_{xx}$ to rather a broad plateau and a more predominant first peak. The two-plateau shape of the out-of-plane coefficient is preserved.
 The transmission spectra are measured with an incidence angle perpendicular the sample surface. The vector of the electric field is thus almost parallel to the sample surface and therefore also yielding mainly an in-plane contribution.  
The overall shape of the measured spectrum, in particular the two-plateau feature, as well as the decrease in absorption with strain is in line with the simulation. We note that a direct comparison between the calculated and measured spectrum should be taken with caution as the theoretical spectrum is performed for the strained ordered bulk phase, whereas in the measured samples SCIO is grown as a thin film on different substrates with a SrTiO$_3$ capping layer. Thus interfacial effects both to the substrate and  the capping layer as well as effects of cation disorder, that are not explicitly considered in the calculation, potentially lead to a reduced band gap and an early onset of the measured spectrum, while the first peak in the simulation lies at higher energies. On the other hand, the $G_{0}W_{0}$+BSE spectrum renders a good agreement concerning the position of the second peak/plateau at $\sim 2.6$ eV, signifying the importance of many body effects. 





\section{\label{sec:Summary}Conclusion}

  Based on density functional theory calculations taking into account a Hubbard $U$ term and spin-orbit coupling, we present a comprehensive investigation of the structural, electronic, magnetic and optical properties of the double perovskite SCIO in comparison to the end members SCO and SIO. While SCO is a ferromagnetic metal and SIO a nonmagnetic semimetal, SCIO is an antiferromagnetic insulator with a Co spin and orbital moment of $\sim 2.4$ and 0.3 $\mu_{B}$, respectively, and most remarkably, a significant spin and orbital moment of Ir of $\sim 1.2$ and 0.13~\mub. Analysis of the orbital occupation w.r.t the end members confirms a considerable electronic reconstruction and a charge transfer from the Ir minority to majority spin states, rendering a substantial magnetic moment at Ir, that explains previous experimental findings~\cite{Tjeng}, as well as $\sim 0.15 e^-$ from Ir to Co, corresponding to an Ir$^{4+\delta}$, Co$^{4-\delta}$ configuration, pointing towards a $j_{\rm eff}=1/2$ Mott insulating state, similar to other iridates. 
	
At the Co sites, the changes in occupation have opposite trends for \eg\ (depletion) and \tg\ states (enhancement) and the latter show orbital and strain-dependence with increase of the \dxy\ occupation by 0.28 to 0.38$e^-$ from compressive to tensile strain versus 0.19 to 0.14$e^-$ (\dxz) and  0.15 to 0.11 (\dyz). The band gap also changes nonmonotonically with biaxial strain from 163~meV (144 meV with SOC) at $a_{\rm NGO}$, to 235~meV (189~meV with SOC) at $a_{\rm STO}$ and, finally, 187 meV (151 meV with SOC) $a_{\rm GSO}$. Quasiparticle effects enhance the band gap by $\sim0.34$ eV with a subsequent slight redshift due to excitonic contributions.  

The absorption coefficient obtained within the  I.P. approximation indicates a two broad peaks/plateaus shape at $\sim 1.2$ and $2.0-2.3$ eV, the first one dominates in particular in the in-plane component of the dielectric constant. The relative differences between the first and the second peak and between in- and out-of-plane contributions is reduced from compressive to tensile strain. Quasiparticle effects within $G_{0}W_{0}$ blueshift the spectrum, while excitonic effects redistribute spectral weight to lower energies. The  two-plateau feature is also observed in the transmission measurements on epitaxial films grown via MAD on LSAT, STO and GSO substrates. While the early onset of the experimental spectrum implies interfacial effects and possible contribution of cation disorder, many body effect improve the agreement to experiment concerning the position of the second peak at $\sim 2.6$ eV.

Overall, SCIO is an intriguing example how the combination of $3d$ and $5d$ elements in a double perovskite allows to achieve electronic phases that are not available in the end members and thus points towards a promising path to tailor exotic materials as an alternative to oxide interfaces~\cite{Lorenz_2016}.

\begin{acknowledgments}
  We acknowledge funding by the Deutsche Forschungsgemeinschaft (DFG, German Research Foundation) within CRC/TRR80 (Project number 107745057), project G03 and CRC1242 (project number 278162697, subproject C02 and A01) as well as computational time at the Center for Computational Sciences and Simulation of the University of Duisburg-Essen on the supercomputer magnitUDE (DFG grants INST 20876/209-1 FUGG, INST 20876/243-1 FUGG) and Leibniz Rechenzentrum, project pr87ro. We also would like to thank S. Moss for performing initial calculations with Wien2k.
\end{acknowledgments}


%

\end{document}